\documentclass[acsnano, twocolumn, superscriptaddress,showpacs]{revtex4}

\usepackage{amsmath}
\usepackage{amstext}
\usepackage{graphicx}
\usepackage{rotating}
\usepackage{tabularx}
\usepackage{color}
\usepackage[english]{babel}
\usepackage{times}

\def\CS{Cu$_2$OSeO$_3$ }
\def\BS{Bi$_2$Se$_3$ }

\graphicspath{{figs/}}

\begin{document}

\title{High Figure of Merit Magneto Optics from Interfacial Skyrmions on Topological Insulators}

\author{Tonmoy K. Bhowmick}
\thanks{tbhow001@ucr.edu, These authors contributed equally}
\affiliation{Department of Electrical Engineering, University of California - Riverside, CA 92521}
\author{Amrit De}
\thanks{amritde@gmail.com, These authors contributed equally}
\affiliation{Department of Electrical Engineering, University of California - Riverside, CA 92521}
\author{Roger K. Lake}
\thanks{rlake@ece.ucr.edu}
\affiliation{Department of Electrical Engineering, University of California - Riverside, CA 92521}

\date{\today}

\begin{abstract}
In the Kerr rotation geometry, magneto optic memory devices typically suffer from low
figure-of-merit (FOM) and long write times.
We show that skyrmions formed at the interface of a thin-film multiferroic and a topological insulator can give rise to high FOM magneto optic Kerr effects (MOKEs).
Huge differential MOKE can arise in parts of the phase diagram.
Resonance like features in the MOKE spectra arising from the induced low energy TI bandgap,
the multiferroic-film thickness, and the high energy Drude like behavior are resolved and explained.
The Fermi level dependence of the MOKE signatures is distinct for the different magnetic textures.
This has broad implications for magnetic texture characterization, electro-optic modulators and isolators and  high density magnetic optic memory.
\end{abstract}

\maketitle
\date{\today}

%==============================
%\section{Introduction}
%==========================================================

Magneto-optic (MO) phenomenon such as Kerr and Faraday rotations, Voigt effect, magneto-plasma
reflections and cyclotron resonances arise from broken time reversal symmetry.
Some MO effects also arise from the electronic structure's topology,
and hence are used to study quantum hall effects\cite{Kukushkin1996},
Kerr rotations in topological insulators~(TIs)\cite{Tse2010,Tse2011},
magneto-electric optical effects\cite{Arima2008,atmatzakis2018magneto},
magneto-chiral effects\cite{Sessoli2015},
and skyrmions and their Hall effects\cite{Vomir2016,woo2018current,Jiang2017}.

%Chiral topological structures generate their own MO signatures \cite{Kukushkin1996,Tse2010,Tse2011,Arima2008} such as magneto-chiral
%effects \cite{Sessoli2015} and skyrmion Hall effects\cite{Vomir2016,Jiang2016}.

The magneto optic Kerr effect (MOKE) in the polar configuration is
particularly interesting due to its application in optical reading--out
of magnetically stored information\cite{Hansen1990,Suzuki1992,Mansuripur2000}
A MOKE device can be characterized by its figure-of-merit (FOM)\cite{Mansuripur1986,Challener1995,Taussig2008prb}, which usually depends on the Kerr rotation angle and the reflectivity.
Typically, during the MO memory write process, a focussed laser heats the magnetic material to its Curie
temperature, which allows the local magnetic polarization to be flipped.
However, the thermally--assisted write processes can be relatively slow.
For readout, the Kerr rotation is barely one degree for most MO recording materials
\cite{Lairson1993apl,Van1983apl,Egashira1974jap,Reim1988apl},
%(with some notable exceptions such as CeSb\cite{Pittini1996prl}).
which can result in higher readout error rates.
The Kerr rotation can be enhanced by the use of cavity like resonance conditions, but this usually lowers the reflectivity and makes
the memory write process difficult.
%
%Overall MO memories have fallen into disuse in recent times.
%
%
%such as PtFe, PtCo, \cite{Lairson1993apl} PtMnSb\cite{Van1983apl}, MnBi \cite{Egashira1974jap}, TbFeCo, TbFeCoCu\cite{Reim1988apl}.
%However in some cases, such as for  CeSb\cite{Pittini1996prl}, drastic MOKE resonance like enhancements have been reported.
%
%One of the problems with generating large Kerr rotations in a semi-infinite medium is that the required magnetic fields and effective cyclotron frequencies can be unreasonably large\cite{De2002,De2003}.
The required magnetic fields and effective cyclotron frequencies for generating large Kerr rotations in a semi-infinite medium can be unreasonably large\cite{De2002,De2003}.
However, skyrmions can give rise to very high emergent fields.
The emergent magnetic fields from a periodic lattice of skyrmions (SkX) can reach up to $4000$ Tesla,
which is two orders of magnitude larger than what can be generated in laboratories.

\begin{figure}
\centering
\includegraphics[width=0.95\columnwidth]{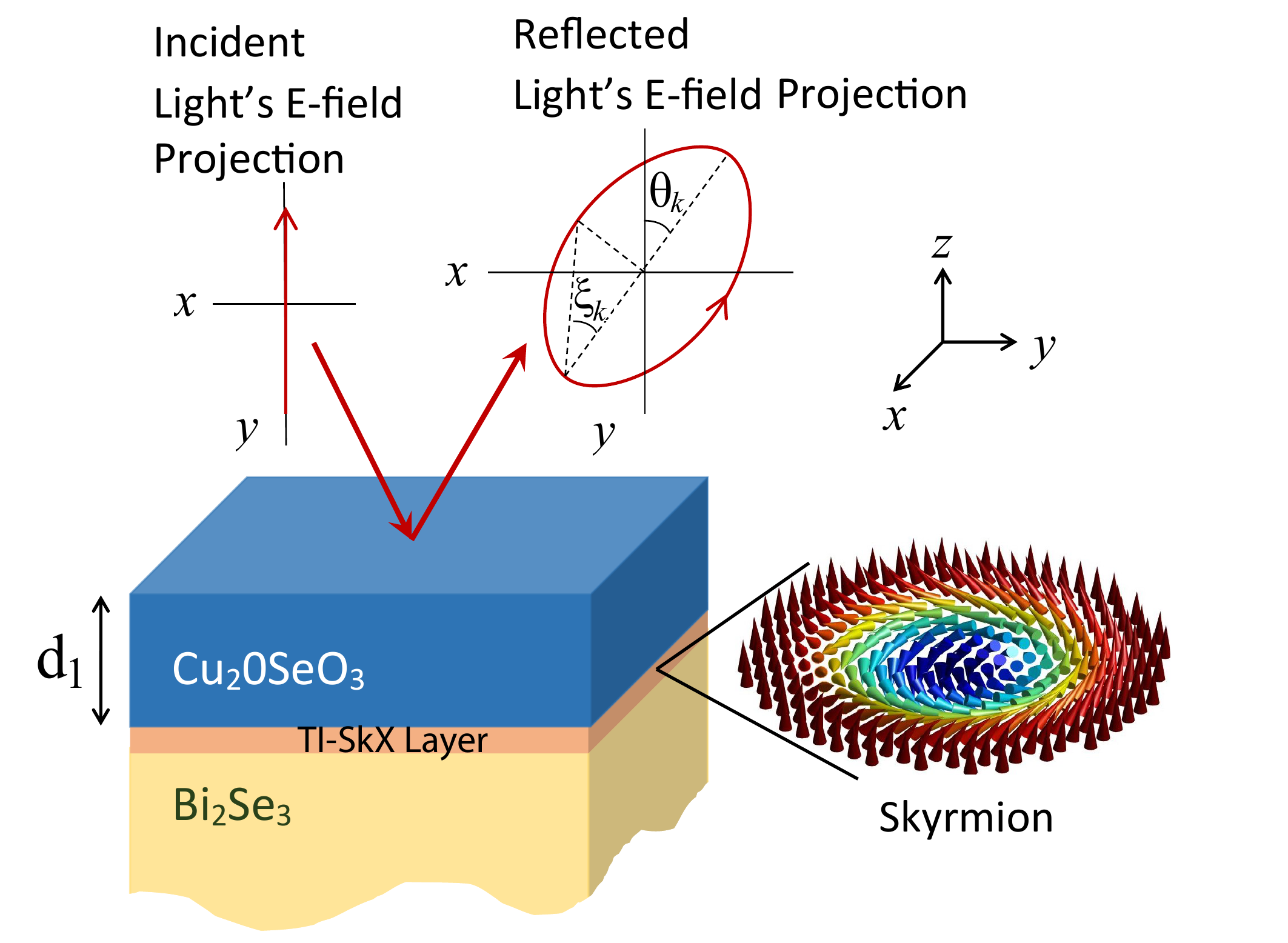}
\caption{MOKE arising from a magneto-optic device with a thin-film B20 material deposited on top of a semi-infinite TI.
A SkX exists at the interface of the TI and B20 compound.}
\label{fig:TF}
\end{figure}

Skyrmions are topological particle-like configurations of continuous vector
fields and are regarded as analogs of magnetic monopoles as each
skyrmion is associated with a quantized magnetic flux\cite{gross2018skyrmion,li2014tailoring}.
Their non-trivial topology is a result of competing Dzyaloshinkii-Moriya-(DM), Heisenberg- and Zeeman-interactions
\cite{Jiang2017,tolley2018room,luo2018reconfigurable}.
In addition, Skyrmions represent a new type of magnetic order\cite{rossler2006spontaneous,muhlbauer2009skyrmion},
and they have been observed in B20
compounds such as MnSi\cite{Tonomura2012ACS,Yu2013ACS}, FeCoSi\cite{Yu2010Nat},
FeGe\cite{Huang2012PRL} and in {\CS} which is a multiferroic\cite{Seki2012Sci}.
Skyrmions have been proposed for use in novel high density electrically controllable racetrack
memories\cite{zhu2018skyrmion,yu2016room}.

The use of magnetic\cite{Mochizuki2012PRL} and electric fields\cite{Mochizuki2015APL}
has been proposed to switch between the topologically nontrivial skyrmion spin texture
and the topologically trivial ferromagnetic spin texture.
Experimentally it is possible create and erase individual skyrmions using spin
polarized currents\cite{Romming2013Science}.
Also magnetic skyrmions can be electrically created on a thin-film of
a chiral-lattice magnetic insulator within a few nanoseconds by
applying an electric field which couples to the noncollinear skyrmion
spins\cite{nakatani2016electric, white2012electric, koshibae2015memory,
schott2017skyrmion, okamura2016transition,mochizuki2015writing, romming2013writing}.
In Cu$_2$OSeO$_3$, the noncollinear skyrmion spin structure in the host material behaves
like a multiferroic due to spin–orbit coupling.
This enables skyrmion manipulaton via electric fields instead of electric
currents \cite{everschor2011current, zang2011dynamics} or heat pulses \cite{koshibae2014creation}.
More recently it has been shown that one can reversibly switch between
topologically inequivalent ferromagnetic phases and skyrmion phases
using electric fields \cite{Hsu2016Nat}.
A topological charge analysis of skyrmion dynamics, energetics, creation, and stability
was recently described \cite{yin2016topological}.

Skyrmions combined with other topological materials can lead to emergent functionalities.
In this paper, we show that
a skyrmion lattice at the interface of a thin-film B20 compound (such as Cu$_{2}$OSeO$_{3}$)
and a semiinfinite TI (e.g. Bi$_{2}$Se$_{3}$)
gives rise to a high FOM-MOKE (see Fig. \ref{fig:TF}).
The large MOKE can be exploited for characterization of the spin texture,
or application in a magneto--optic memory device.
%

%The skyrmion spin texture proximity coupled to a TI surface becomes charged
%due to the localized Dirac surface states spatially confined inside
%the radius of the skyrmion\cite{hurst2015charged}.
%%
%The TI stabilizes the skyrmion because of the peculiar characteristics of the surface states, which also
%provides an electronic handle to manipulate skyrmions.
%%
%Experimentally, the formation of skyrmions on the surface of a TI has been observed~\cite{yasuda2016geometric}.

Skyrmions can form on a TI's surface\cite{yasuda2016geometric} and can become charged\cite{hurst2015charged}.
The B20--TI heterostructure provides
electric field switching of the skyrmion spin texture and a large MOKE
from the skyrmion phase.
The large differential MOKE leads to fast and high fidelity magneto--optical bit readout.
Electric field switching provides fast, low--power writing.
The emergent properties of the B20--TI heterostructure provide the physical mechanisms
for facile reading and writing of information bits in a topological magneto-optic memory.

For simplicity we consider a device geometry where a seminfinite TI is capped by a thin {\CS} layer as shown in Fig. \ref{fig:TF}.
We consider interfaces with a ferromagnetic (FM) texture along with N\'{e}el- and Bloch type skyrmions.
This simple geometry allows us to best highlight the main MOKE spectral features manifesting from different physical effects.
These are: the low energy topological MOKE, thin-film induced enhancement of MOKE and MOKE occurring at the high energy plasma frequency.

%========================================================================================
%\section{Effective Models}
%========================================================================================
\emph{Model and Method:}
We begin by constructing a model where the surface of the TI is coupled to the background spin texture of a SkX.
A low energy effective model describing the surface state of Bi$_2$Se$_3$ is used where the
surface state consists of a single Dirac cone at the $\Gamma$ point \cite{lu2010massive}.

We consider the low energy effective Hamiltonian\cite{xia2009observation}  describing the decoupled top and bottom surface states of a TI %is as follows:
%\begin{equation}
%H = \hbar v_{F} (k_{x}\sigma_{y}- k_{y}\sigma_{x})
%\end{equation}
%
To avoid the well known Fermion doubling problem on discrete tight-binding lattice, we have added a $k^{2}\sigma_{z}$ term to the Hamiltonian of TI\cite{susskind1977lattice,hong2012modeling}
\begin{equation}
H =  \hbar v_{F} (k_{x}\sigma_{y}- k_{y}\sigma_{x})-  \zeta \hbar v_{F}  (k_{x}^{2}+k_{y}^{2})\sigma_{z},
\end{equation}
where the Fermi velocity $v_{F}$, is a  material constant.
This momentum space Hamiltonian can be transformed into the following real space tight binding Hamiltonian coupled to the Skx spin texture on a rectangular lattice
% by replacing $\mathbf{k}$ with $-i \nabla$ and the differential operators are discretized in a rectangular lattice using finite difference to obtain  lattice Hamiltonian
%
\begin{align}
  H  = \sum_{i}c_{i}^{\dagger} \epsilon c_{i} - \sum_{<i,j>} (c_{i}^{\dagger} t c_{j} + h.c.)
    - J_{H}\sum_{i}c_{i}^{\dagger} \boldsymbol{\sigma}_{i} \cdot \textbf{S}_i c_{i}
\label{eq:H_all}
\end{align}
where ${\boldsymbol{S}}_i$ is the localized spin of the skyrmion on site $i$ which couples to the TI through the Hund's rule coupling constant $J_{H}$.
Here $t$ is the nearest neighbor hopping along the $x$ and $y$
directions, $\boldsymbol{\sigma_{i}}$ is the spin of the itinerant electron, and $\epsilon$ is the on site potential.
Periodic boundary conditions are imposed along both the $x$ and $y$ directed edges of a square unit cell consisting
of $9 \times 9$ lattice sites.
We define two dimensionless parameters $t_1 = \hbar v_F/2a$ and $t_2 = \zeta/a$.
Here, $\epsilon = 8t_1t_2\sigma_{z}$, $t_{x} = -i t_1\sigma_{y} - 2t_1t_2\sigma_{z}$,
$t_{y} = i t_1 \sigma_{x} - 2t_1t_2\sigma_{z}$.
For our numerical simulations we choose the discretization length $a = 15 {\rm \AA}$, $\zeta  = 5 {\rm \AA}$, $v_{F} = 0.5 \times 10^{6}$ m$/$s and $J_{H} = 40$ meV. \cite{yasuda2016geometric}
The interplay between $t_{x,y}$ and $J_H$ dictates the anomalous Hall conductivity of the TI surface in this model.
The magnetization of a single skyrmion can be described by
\begin{equation}{
  {\textbf{n}(\textbf{r})} = \left[\sin\vartheta(r)\cos\varphi(\phi),\sin\vartheta(r)\sin\varphi(\phi),\cos\vartheta(r)\right],
\label{nr}}
\end{equation}
where $\vartheta(r = 0) = 0$, $\vartheta(0< r < R_s) = \pi (1-\frac{r}{R_{s}})$ and
$\vartheta(r>R_{s}) = \pi$. \cite{lado2015quantum}
$\varphi(\phi) = m\phi + \nu$ and $\phi=\tan^{-1}[y/x]$ where the helicity of the skyrmion is defined by the phase $\nu$.
\cite{nagaosa2013topological}
In this paper, we consider both N\'{e}el type  $(m =1,\nu=0)$ and Bloch type skyrmions $(m=1,\nu=\pi/2)$.
The skyrmion is centered in the unit cell, and the
magnetization $\textbf{n}(\textbf{r})$ is evaluated at each lattice site $i$ to obtain $\textbf{S}_i$ in Eq. (\ref{eq:H_all}).

The $z$-component of $\textbf{n}(\textbf{r})$ acts on the electron like position dependent Dirac mass,
while the in-plane components can give rise to an emergent gauge field.
The total magnetic flux enclosed in a unit skyrmion cell is one flux quantum,
$\Phi_{0} = {h}/{e}$, independent of the skyrmion radius, $R_s$. \cite{hamamoto2015quantized}
The effective magnetic field is $B_{\rm eff} = -\Phi_{0}/(4R_{s}^{2})$. \cite{kanazawa2015discretized}
%

%===================================================================================================
%\section{Electronic Structure}
%===================================================================================================
\begin{figure}
  \includegraphics [width=0.95\columnwidth]{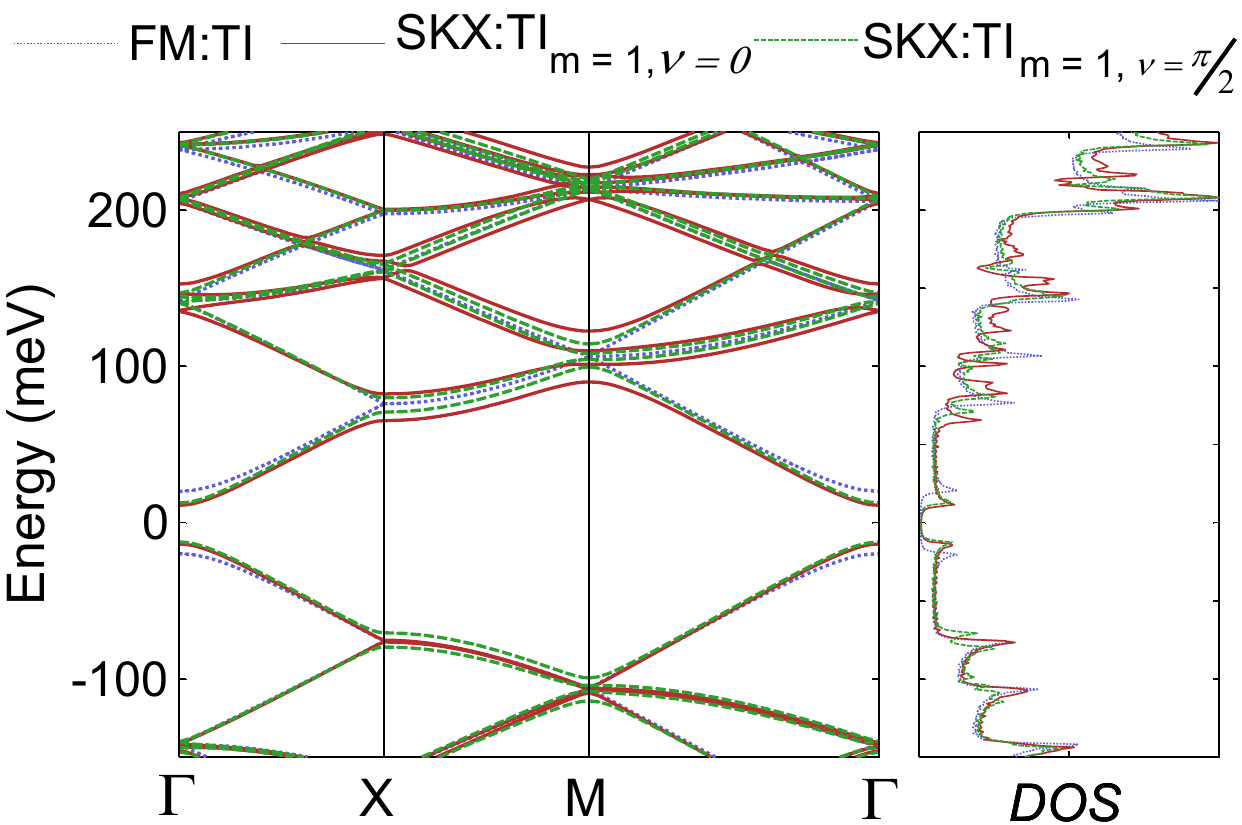}
  \caption{Electronic bandstructure and density of states for proximity coupled FM:TI system and SkX:TI with $m=1$, and $\nu=0$ and $\nu=
\pi/2$}
   \label{fig:band}
\end{figure}

\emph{Electronic Structure:}
The numerically calculated band structure and the corresponding density of states (DOS) are shown in
Fig. \ref{fig:band} for the
magnetic textures proximity coupled to the TI.
This includes the ferromagnetic case (FM:TI),
the N\'{e}el type skyrmion ($m=1,\nu=0$),
and the Bloch type skyrmion ($m=1,\nu=\pi/2$) proximity coupled to the TI's surface state.
%
%In Fig. \ref{fig:band}-(a) it is shown that
The energy gap at $\Gamma$ resulting from proximity coupling to the N\'{e}el and Bloch type
skyrmions is less than the gap resulting from proximity coupling to the FM state.
This can be understood from the fact that the skyrmion creates a hole in the background FM
texture that reduces the total $\hat{z}$ component of the magnetic moment.
The presence of the skyrmions also breaks the degeneracy in other higher energy regions of the spectrum.
%

%============= dielectric tensor calculations =============================================
\emph{Dielectric Tensor Components:}
%==========================================================================================
Observables such as magneto optic effects and quantum Hall type phenomenon manifest
themselves through the dielectric tensor components
which depend on the electronic structure and topological properties.
In the case of normal optical incidence for an SkX on a TI,
the magnetization is along $z$, which is perpendicular to the surface and parallel
to the direction of light propagation, similar to the polar Kerr effect.
The $x$ and $y$ directions preserve in-plane symmetry.  The complex $3\times 3$ dielectric tensor has
diagonal $[\epsilon_{xx},\epsilon_{yy},\epsilon_{zz}]$ terms and the off-diagonal $\epsilon_{xy}$ terms which are topology dependent.

The matrix elements of the optical conductivity tensor can be obtained from the band structure calculations using the Kubo formula
\cite{thouless1982quantized, kohmoto1985topological} as follows:
\begin{widetext}
\begin{eqnarray}{
\sigma_{\i\j} =  \frac{ie^2}{(2\pi L)^2\hbar} \displaystyle\int\frac{d\boldsymbol{k}}{2\pi} \sum_{n,l}
\frac{ f_{nl}(\boldsymbol{k}) }{\omega_{nl}(\boldsymbol{k})}
\left(\frac{  \Pi(\boldsymbol{k})_{nl}^{\i}\Pi(\boldsymbol{k})_{ln}^{\j} }{ \omega_{nl}(\boldsymbol{k}) - \omega + i\gamma }
+\frac{  \Pi(\boldsymbol{k})_{ln}^{\i}\Pi(\boldsymbol{k})_{nl}^{\j} }{ \omega_{nl}(\boldsymbol{k}) + \omega + i\gamma }\right)
\label{Kubo}}
\end{eqnarray}
\end{widetext}
where $\Pi^{\i}_{nl}(\boldsymbol{k}) = \langle\psi_n(\boldsymbol{k})|v_{\i}|\psi_l(\boldsymbol{k})\rangle $ is the velocity operator.
Here $\{\i,\j\}\in\{x,y\}$, $\gamma$ is a broadening parameter and $\hbar \omega_{nl}(\boldsymbol{k}) = E_{n}(\boldsymbol{k}) - E_{l}(\boldsymbol{k})$, is the energy difference of an optical transition between an unoccupied band, $n$ and an occupied band, $l$.
$f_{nl}(\boldsymbol{k})=f_n(\boldsymbol{k})-f_l(\boldsymbol{k})$, where $f_n(\boldsymbol{k})$ is the Fermi filling factor.

%---------- kappa plus bulk bands issue ----------------------------------------------
Since we are using an effective Hamiltonian to obtain $\sigma_{\i\j}$, we compensate for the missing higher band contributions in Eq. (\ref{Kubo}) by adding a $\kappa/(\omega+i\gamma$) term while relating the optical dielectric tensor to the conductivity tensor.
\begin{eqnarray}
\epsilon_{\i\j}(\omega) = \varepsilon_o\delta_{\i\j} - \frac{4\pi i}{\omega}\sigma_{\i\j} -\frac{\kappa}{\omega + i\gamma}
\end{eqnarray}
 Here $\kappa$ is adjusted so that the relative zero frequency dielectric constant $\epsilon_0$ matches the known experimental value. $\varepsilon_o$ is the vacuum permittivity.
For Bi$_2$Se$_3$, $\epsilon(\omega)$ is given by Lorentz oscillator fits to experiment \cite{Eddrief2016}, which in the low energy regime ($0-30$ meV) is essentially a constant.
In the effective surface model, the bulk band contributions to the momentum matrix elements in $\epsilon_{\i\i}(\omega)$ are not included.
However, this does not change the qualitative behavior of the effects shown in this paper, since the calculated $\epsilon_{\i\i}(\omega)$s are qualitatively similar to experiment\cite{Eddrief2016}.
Therefore we argue that all the higher energy MOKE features disccused in this paper would still be seen in experiments, but at higher optical frequencies.

%---------- Berry Connection ---------------------------------------------------------
%
The dielectric function consists of Berry connection type terms,
$a_{\i}(\boldsymbol{k}) = -i\langle\psi(\boldsymbol{k})|\nabla_{\i}|\psi(\boldsymbol{k})\rangle$,
which behave like a fictitious momentum space gauge potential or an equivalent magnetic field
$b_{z}(\boldsymbol{k}) = \partial_{k_x}a_{y}(\boldsymbol{k}) - \partial_{k_y} a_{x}(\boldsymbol{k})$.
The MOKE can therefore be viewed as an optical manifestation of the Berry curvature via the $\epsilon_{xy}$ term.
This is similar to charge transport, where the ${xy}$ response is proportional to the Chern number
\cite{rossler2006spontaneous, jackson1985skyrmion, munzer2010skyrmion, yu2012skyrmion},
which is the integral of Berry curvature over the
first Brillouin zone, $\mathcal{C} = \frac{1}{2\pi}\int d^{2}k b_{z}(\boldsymbol{k})$.
There are differences between MOKE and quantum Hall effect type topological manifestations due to conduction $\leftrightarrow$
valance transitions and the frequency dependence in optics.
$\epsilon_{xy}$ depends on the topological charge across a gap.
%In our case the first electron(hole) band contributes a Chern number of $-1(+1)$ which constitute the effective topological charge for
%combined TI-SkX system.
%
Usually $\epsilon_{xy}$ is much smaller than the diagonal $\epsilon_{xx}$.
In our case the three different magnetic textures
result in the same Chern number across the fundamental gap of the proximity induced magnetic TI surface state,
so that $\epsilon_{xy}$ does not vary much when the Fermi energy is set to 0.

%--------------------------------------------------------------------------------------
%------------------------------ MOKE ----------------------------------------------
\emph{Magneto-Optics:}
The complex in-plane index of refraction is ${n}_\pm = \sqrt{\epsilon_\pm}=\sqrt{\epsilon_{xx}\pm i\epsilon_{xy}}$ where, the $+(-)$ signs represents right(left)
circularly polarized (RCP(LCP)) light propagation.
The complex MOKE effect is expressed as:
$\Theta_k = \mathcal{\theta}_k +i\xi_k$
where the Kerr rotation and ellipticity are, respectively
\begin{eqnarray}
\theta_k &=& (\Delta_+-\Delta_-)/2 \\
\xi_k &=& (|r_+|-|r_-|)/(|r_+|+|r_-|) .
\end{eqnarray}
Since the eigen-modes here are LCP and RCP, the Kerr rotation angle can be expressed as the phase difference between these two modes.
The complex phase $\Delta_{\pm}$ can be obtained from the Fresnel reflection coefficients $r_\pm$.

%It should be noted that the commonly used expression for calculating MOKE,
%$\Theta_k  \approx {\epsilon_{xy}}/{(1-\epsilon_{xx})\sqrt{\epsilon_{xx}}}$, can only be used when $\epsilon_{xx} \gg \epsilon_{xy}$ and for
%small Kerr rotations only \cite{Argyres1955,De2002}.
%%
%MOKE of skyrmions cannot be treated using this expression since skyrmions have huge
%emergent magnetic fields of the order of 1000's of T.
%%
%These large effective fields cause large reflection edge splittings,
%the effect of which on MOKE are not captured by this approximation.

The MOKE arising because of the thin-film structure can be significantly altered
by internal reflection at various interfaces of the layers.
A $2\times2$ characteristic matrix method can be used to characterize the MOKE spectra of a multilayer structure
at normal incidence, assuming that the materials are homogeneous and isotropic.
The transfer matrices are in the LCP/RCP eigenmode basis, which for $N$ parallel layers is:
\begin{eqnarray}\label{TM}
\boldsymbol{S}^\pm =
\prod_{j=0}^{N}
\frac{1}{t_{j,j+1}^\pm}
  \left[
\begin{array}{cc}
  r_{j,j+1}^{\pm} &  1 \\
  1 &  r_{j,j+1}^{\pm} \\
\end{array}
\right]
  \left[
\begin{array}{cc}
  e^{i\beta_{j+1}^\pm} & 0 \\
  0 &  e^{-i\beta_{j+1}^\pm} \\
\end{array}
\right]
\end{eqnarray}
The Fresnel reflection and transmission coefficients at normal incidence for each interface are respectively given by:
$r_{j,j+1}^\pm = (n_j^\pm-n_{j+1}^\pm)/(n_{j}^\pm + n_{j+1}^\pm)$ and
$t_{j,j+1}^\pm = (2 n_j^\pm)/(n_{j}^\pm + n_{j+1}^\pm)$
The phase factor is given by $\beta_j = (2\pi/\lambda)n_j d_j$,  where $d_j$ is the thickness of the $j^{th}$ layer and $\lambda$ is the optical
wavelength. The complex reflection coefficient from the resultant characteristic transfer matrix is:
$r^\pm = S_{12}^\pm/S_{11}^\pm = |r^\pm|\exp(-i\Delta_\pm)$ where $S_{\i\j}^\pm\in\boldsymbol{S}^\pm$.
The observed reflective intensity is $R_\pm=|r_\pm|^2$.
%

%-------------------------------------------------------------------------------------------

\begin{figure}
  \includegraphics [width=1\columnwidth]{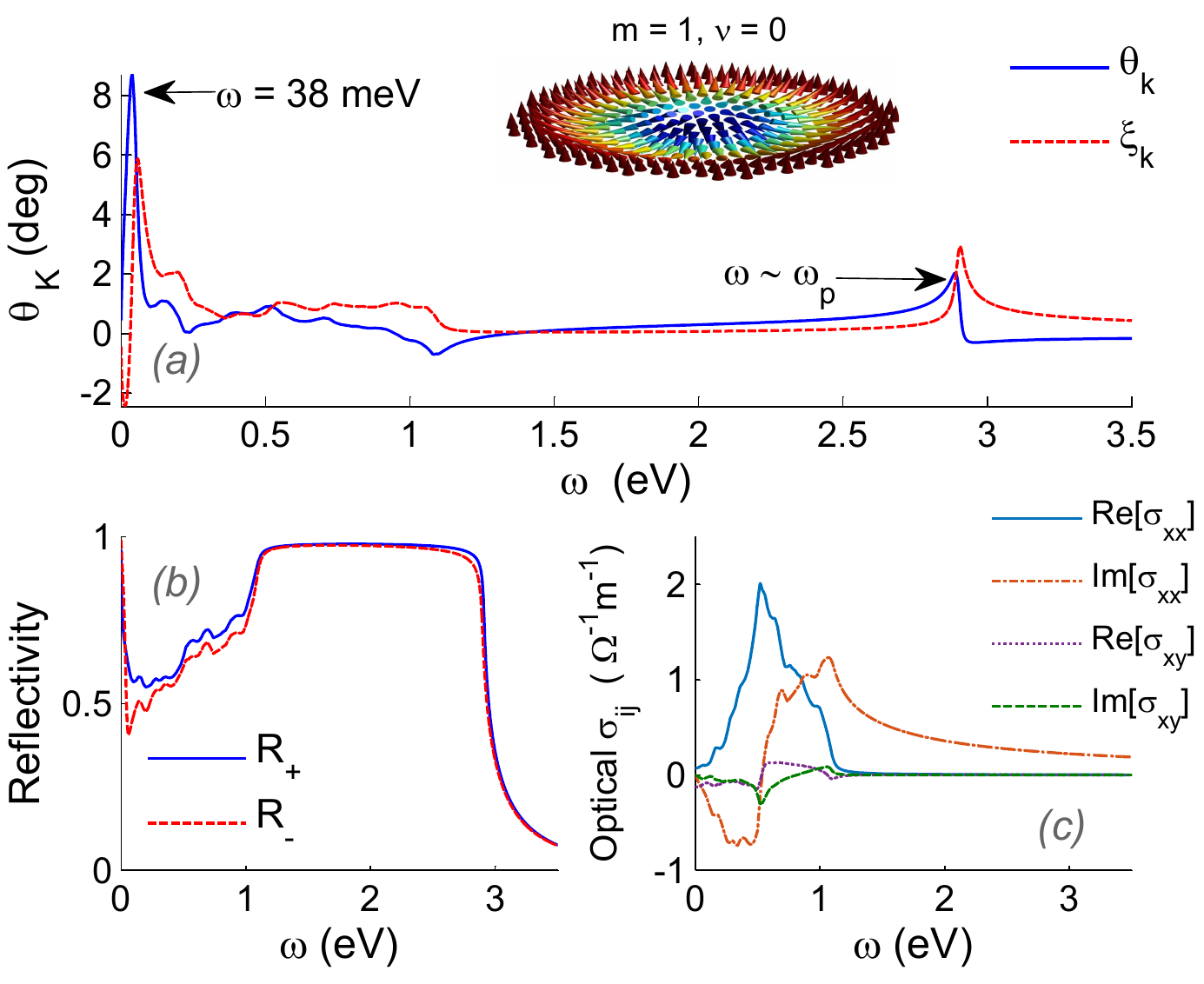}
  \caption{ (a) Kerr rotation and ellipticity for a single SkX:TI interface for N\'{e}el type skyrmion $m=1, \nu=0$ (b)
      Reflectivity for left- and right-circularly-polarized light. (c) The corresponding optical conductivity of SkX:TI system}
   \label{fig:MOKEsi}
\end{figure}

%===================================================================================================
%\section{Discussion}
%===================================================================================================
%%============================================================
%\subsection{Effects from a Single Interface}
%%============================================================
\emph{Discussion:}
We first consider only an ideal single interface for our initial analysis.
The MOKE spectra and reflectivity and the optical dielectric function are shown for a
N\'{e}el type skyrmion ($m=1,\nu=0$) on a TI in Fig. \ref{fig:MOKEsi} for illustration.
The corresponding spectra for FM:TI and Bloch type skyrmion ($m=1,\nu=\pi/2$) on a TI look very similar.

Two distinct resonance like features can be seen for the Kerr rotation and the elipticity.
From the approximation\cite{Argyres1955}: $\Theta_k\approx[n_+-n_-]/[n_+n_--1]$,
it is easy to see that MOKE resonances spectrally occurs whenever
$\epsilon_+\epsilon_-\sim 1$\cite{De2002}.
In Fig. \ref{fig:MOKEsi}, one MOKE resonance is in the low energy regime and occurs at 38 meV.
This is close to the size of the fundamental gap that the magnetic texture
(N\'{e}el type skyrmion in this case) induces in the TI.
We will explore this regime, where the quantization effects occurs, in more detail later in this paper.

%--------------------- Drude Regime ----------------------------------------------------------------
The high energy MOKE resonance features occur in the energy regime where free electron like behavior dominates.
Hence this can be qualitatively understood from the semiclassical Drude model:
$\epsilon_\pm \propto \left( 1- {\omega_p^2}/[{ \omega(\omega \pm \omega_c + i\gamma)}] \right)$,
%
%\begin{eqnarray}{
%\epsilon_\pm = \epsilon_\infty \left( 1- \frac{\omega_p^2}{ \omega(\omega \pm \omega_c + i\gamma)} \right)
%\label{Drude}
%}\end{eqnarray}
where $\omega_p$ is the plasma frequency and $\omega_c$ is the cyclotron frequency.
%Typically if $\omega_c\ll\omega_p$, $R_\pm$ will show a resonance like feature near $\omega=\omega_p$.

If $\epsilon_{xy}\ll\epsilon_{xx}$\cite{Argyres1955},
an additional approximation can be made for the Kerr rotation,
$\Theta_k  \approx {\epsilon_{xy}}/{(1-\epsilon_{xx})\sqrt{\epsilon_{xx}}}$.
From this most commonly used expression for calculating MOKE, it can be seen that MOKE resonances should occur
when $\epsilon_{xx}=1$.
This is also where $R_\pm$ goes to zero and the MOKE resonance occurs near $\omega=\omega_p$\cite{Feil1987,De2002a,Abe2004}.
Note that similarly, magneto-plasmons in graphene give large magneto optic effects\cite{Ferreira2012,Tymchenko2013}.
Using the Drude model, one can derive a simple expression for the spectral occurrence of this MOKE resonance\cite{De2002}
\begin{eqnarray}
%\omega\approx \sqrt{ (\omega_c/2)^2 \pm \omega_p^2 }\\
\omega_\pm\approx\frac{1}{2}\left( \gamma + \sqrt{ \omega_c^2 -\gamma^2 + 4\omega_p^2 } \pm \omega_c \right)
\label{Eq:wmax}
\end{eqnarray}

In an attempt to explain the high energy spectral features using the Drude model, we extract the effective $\omega_p$
and effective $\omega_c$ using the optical sum rules:
\begin{eqnarray}
\displaystyle\int_0^\infty \omega Im[\epsilon_{xx}(\omega)]d\omega &=& \frac{\pi}{2}\omega^2_p\\
\displaystyle\int_0^\infty \omega^2 Re[\epsilon_{xy}(\omega)]d\omega &=& -\frac{\pi}{2}\omega^2_p\omega_c
\label{sumr}
\end{eqnarray}
These values can be substituted into Eq. (\ref{Eq:wmax}) to obtain $\omega_+$, which is listed in Table \ref{tab:list} along with other parameters. The analytically obtained $\omega_+$ is in excellent agreement with $\omega_\theta$ -- which is the frequency
at which the high energy MOKE resonance occurs.
In this high energy regime, the magnitude of $\theta_k^{max}\propto\omega_c\omega_p^2\propto\epsilon_{xy}$ as shown in Table \ref{tab:list}.
Overall if $\omega_c\ll\omega_p$ then the Kerr rotation resonance is at $\omega\approx\omega_p$. Whereas if $\omega_c\gtrsim\omega_p$ then a large MOKE resonance edge split occurs at frequencies determined by $\epsilon_+\epsilon_-\approx 1$\cite{De2002}.
Note that these calculations are for a single interface (i.e. assuming both the \CS and \BS layers are semi-infinite).

\begin{center}
\begin{table}
\renewcommand{\arraystretch}{1}
\begin{tabular}{c|c |c |c |c |c}
 \hline
 \hline
Spin Texture & $\theta_k^{\omega_p}$ & $\omega_p$ (eV)& $\omega_c$ (eV)& $\omega_{\theta}$ (eV)& $\omega_+$ (eV)  \\
\hline
FM           &  2.175   & 2.822 &  0.014 & 2.894 & 2.878 \\ %2.829 -- ignoring gamma
$m= 1, \nu=0$  &   2.032   & 2.816 &  0.013 & 2.888 & 2.872 \\  % 2.828 -- ignoring gamma
$m=1, \nu=\pi/2$  &  2.029   & 2.816 & 0.013 & 2.889 & 2.872 \\  % 2.828 -- ignoring gamma
\hline
\end{tabular}
\caption {
List of high energy ( $>$ 2 eV) Kerr rotation angles and related parameters
due to free electron like behavior  for various magnetic textures on TI.
Maximum Kerr rotation angle $\theta_k^{\omega_p}$, effective plasma frequency $\theta_k^{\omega_p}$,
cyclotron frequency $\omega_c$,
frequency of maximum Kerr rotation $\omega_{\theta}$, and its predicted value $\omega_+$.}
\label{tab:list}
\end{table}
\end{center}

%\begin{center}
%\begin{table}
%\renewcommand{\arraystretch}{1}
%\begin{tabular}{c|c |c}
% \hline
% \hline
%Texture on TI & $\omega_p$ (meV)& $\omega_c$ (meV)  \\
%\hline
%$m= 1, \nu=0$  &  16.1 & 36.9 \\
%$m=1, \nu=\pi/2$ &   15.5 & 144 \\
%\hline
%\end{tabular}
%\caption {List of the effective plasma and cyclotron frequencies for various skyrmion textures on TI.}
%\label{tab:list}
%\end{table}
%\end{center}

%Even though the $\omega_c$ for the FM:TI is much larger than that of $m=1, \nu=0$ SkX:TI, the Kerr rotation is not.
%$\theta_k^{max}$ for the FM:TI is smaller than that for the SkX:TI, where the $m= 1, \nu=\pi/2 $ gives the largest Kerr rotations and has the largest $\omega_c$.
%For the N\'{e}el type skyrmion (where the spins rotate in the vertical plane) the maximum Kerr rotation peaks occur at $\omega\approx
%\sqrt{\omega_p^2 \pm (\omega_c/2)^2}$, which is close to what we can predict from the Drude model.
%Whereas for the Bloch type skyrmion (where the spins rotate out of the vertical plane) its MOKE spectral features can not be explained by the
%Drude model.

\begin{figure}
\centering
\includegraphics[width=0.95\columnwidth]{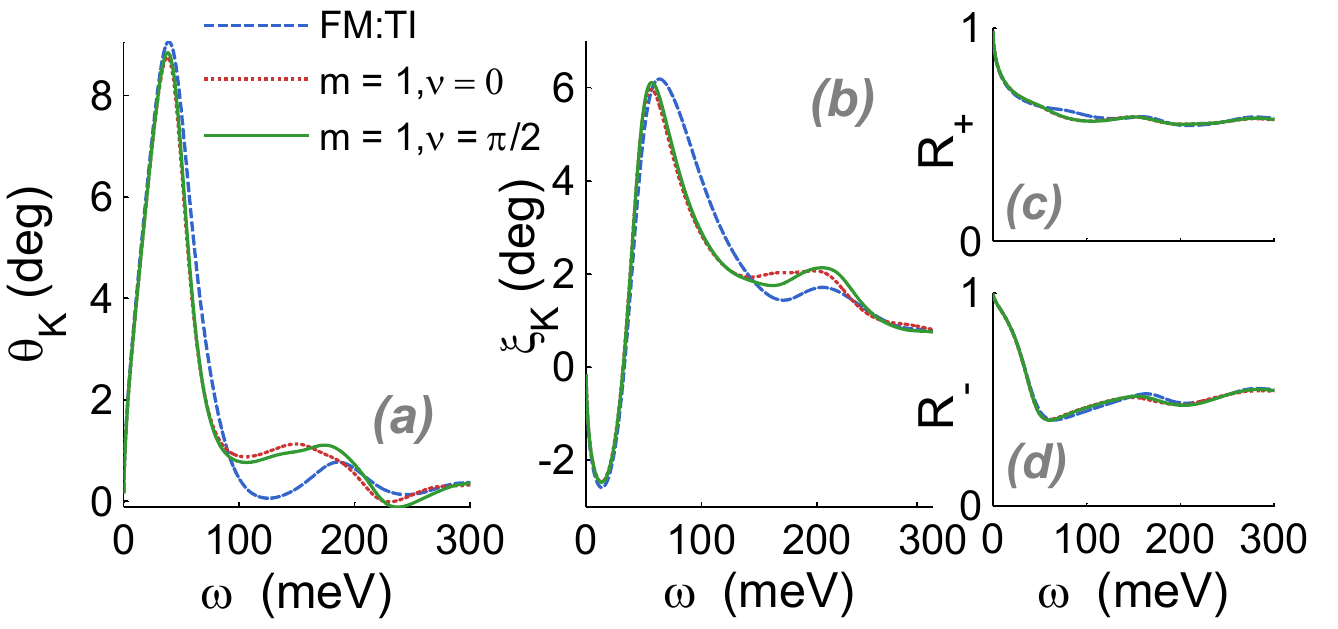}
\caption{ A comparison of the (a) Kerr rotation, (b) ellipticity, (c) RCP reflectivity, and (d) LCP reflectivity
for FM:TI and SkX:TI with m = 1, $\nu=0$ and m = 1, $\nu=\pi/2$.}
\label{fig:SIlow}
\end{figure}

%-------------------------------------------------------------------------------------------------
% Single Interface Low energy regime
%------------------------------------
We next consider the MOKE features around 38 meV.
The MOKE spectra is shown in Fig. \ref{fig:SIlow} in the energy regime below 300 meV.
The maximum Kerr rotation is about $9^\circ$,
which occurs when $R_\pm$ are the furthest apart as this increases $\Delta_+-\Delta_-$.
The Kerr rotation is sufficiently large for magneto-optic recording applications.
In this case, there is not a clear distinguishing feature between the interface with the
FM and the interface with the two different skyrmion spin textures.
For these calculations, the Fermi Energy, $E_f=0$.
The MOKE spectral features in this energy regime are primarily dominated by the gap in the TI surface state,
which is induced by the magnetic textures.
The Chern number in the gap is the same for all three magnetic textures,
and hence there is no clear distinction in the MOKE features for $E_f=0$.

%-------------------------------------------------------------------------------
%----------------------thin film -----------------------------------------------

MOKE can also be enhanced by the resonance like effects that arise from adjusting the film thickness of different materials.
In order to understand the effects of this for our system,
we consider a thin-film structure as shown in Fig. \ref{fig:TF} where a \CS
film of thickness $d_1$ sits on a semi-infinite Bi$_{2}$Se$_{3}$ layer.
The dispersion relation for \CS was obtained using experimentally fitted Lorentz oscillators\cite{Miller2010prb}.
As is typical with multiferroics, there are several low energy phonon modes present in
the \CS dielectric function which are also reproduced by the model\cite{Miller2010prb}.
We assume that the SkX system exists in a 1 nm thin layer at the interface of the \CS thin film and seminfinite Bi$_{2}$Se$_{3}$ layer.
The SkX:TI effects manifest themselves via $n_\pm$.
The refractive index of air is used for the semi-infinite media above the \CS layer as shown in Fig. \ref{fig:TF}.

\begin{figure}
\centering
\includegraphics[width=1\columnwidth]{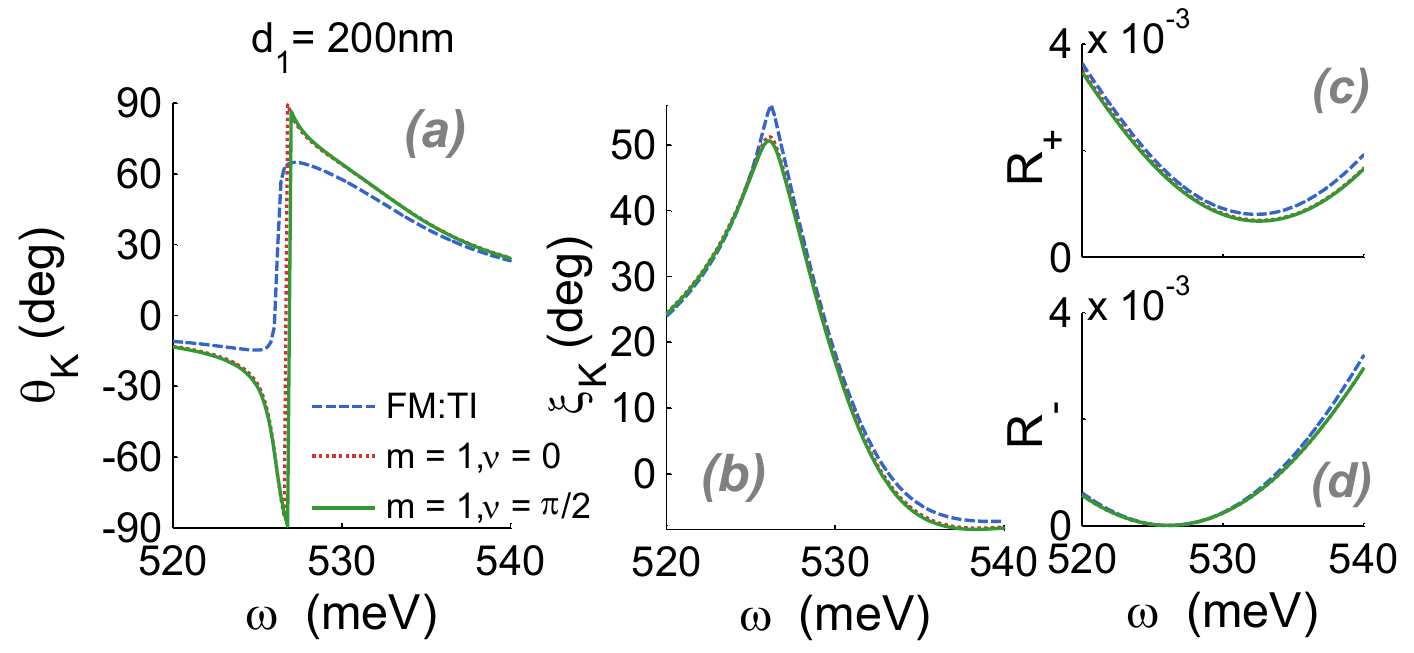}
\caption{A close-up comparison of the $d_1=200$ nm thin-film induced resonances in the (a) Kerr rotation and (b) ellipticity.
The corresponding (c) RCP reflectivity and (d) LCP reflectivity is also shown for the FM:TI- and SkX:TI interfaces with Bloch (m=1, $\nu=\pi/2$) and N{\'e}el (m=1, $\nu=0$) type skyrmions.}
\label{fig:TFres}
\end{figure}

\begin{figure}
\centering
\includegraphics[width=1\columnwidth]{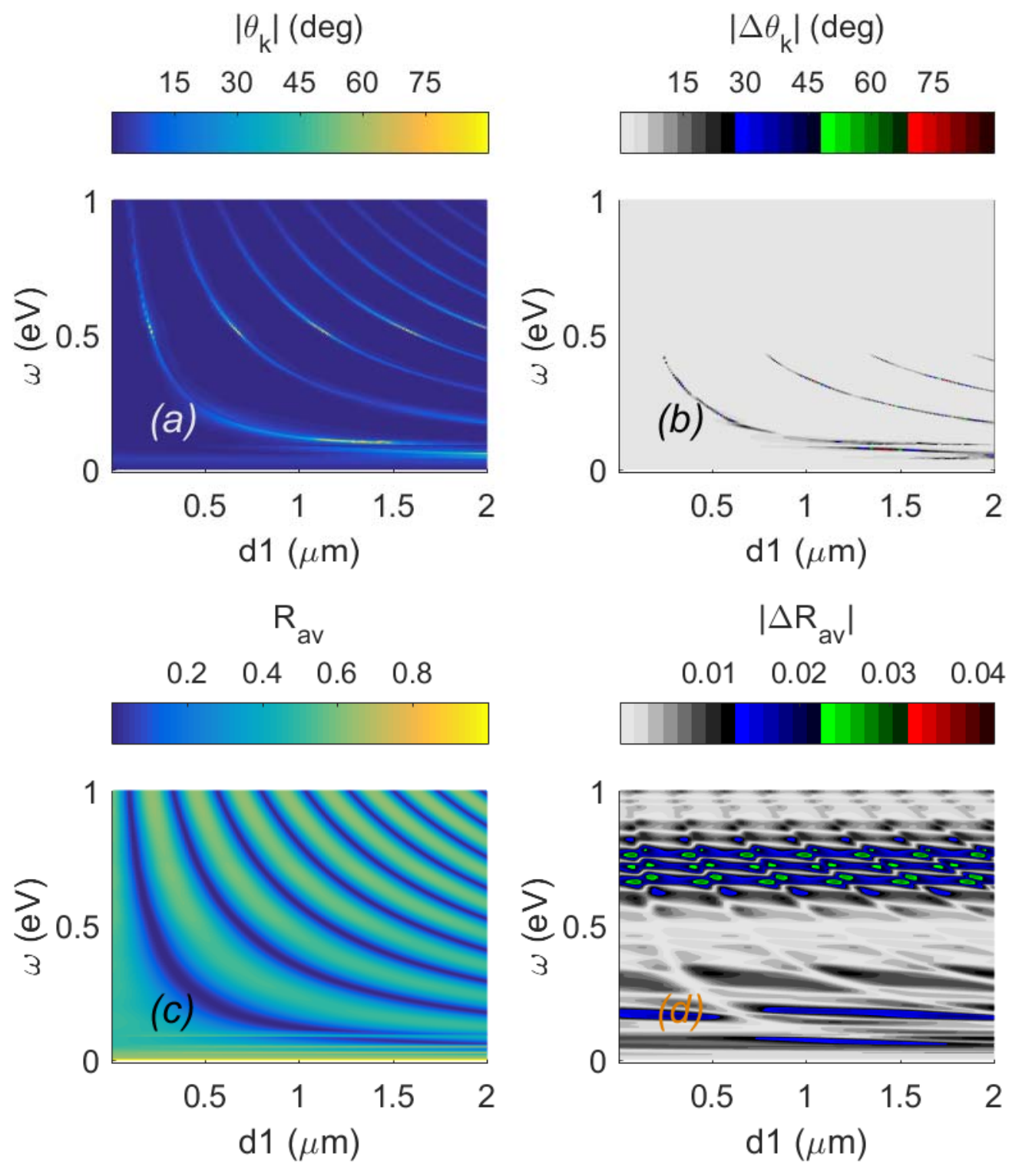}
\caption{(a) The Kerr rotation for the N\'{e}el type ($m=1,~\nu=0$) interface (b) and its Kerr rotation difference
with the FM interface, $\Delta\theta_k=\theta_k^{FM:TI}-\theta_k^{SkX:TI}$
as a function of frequency and \CS film thickness $d_1$.
(c) Average reflectivity $R_{av}=(R_++R_-)/2$ for  N\'{e}el type interface and (d) Difference in the average reflectivity: $\Delta R_{av}=R_{av}^{FM:TI}-R_{av}^{SkX:TI}$}
\label{fig:3D}
\end{figure}

A comparison of the magneto optic Kerr rotation, ellipticity,
LCP and RCP reflectivity spectra are shown in Fig. \ref{fig:TFres} for
a N\'{e}el type skyrmion, the Bloch type skyrmion, and the FM:TI for a \CS
layer with film thickness $d_1=200~\mu m$.
At this particular film thickness, $R_-\rightarrow 0$, which causes a huge Kerr rotation
resonance of $\sim90^\circ$ as the phase difference between LCP and RCP is maximized.
Also for a given film thicknesses for the \CS layer,
the MOKE spectra can be notably different between the different SkX textures and the FM state.
In this example the maximum Kerr rotation arising from the FM:TI is $\sim 60$ deg and occurs at a different optical frequency.

%%***
%The maximum Kerr rotation spectrally occurs in the vicinity of the effective $\epsilon_+\epsilon_-\sim 1$ which can be deduced from the
%approximation: $\Theta_k\approx[n_+-n_-]/[n_+n_--1]$\cite{Argyres1955,De2002}.
%%***

%%==============================================================================
% 3D phase diagram
%%===============================================================================

%We next examine the MOKE phase diagram as a function of $\omega$ and $d_1$ more carefully.
%%

%Skyrmion textures proximity coupled to a TI surface can become charged\cite{hurst2015charged}
%%due to the localized Dirac surface states confined inside
%%the skyrmion
%which provides an electronic handle to manipulate them.
%This could be useful for a MO memory device,
%due to the possibilities of fast reversible electrical switching between a FM- and a skyrmion-phase\cite{Hsu2016Nat}.
%These different topologically inequivalent FM- and SkX phases represent classical bits. Binary-logic such as this should enable high fidelity bit readout using MOKE.
%%The memory read/write process should be fast.

In order to achieve low error magneto-optic readouts,
the different states should be maximally discernable.
We therefore examine the phase diagram, as a function of $\omega$ and $d_1$, for the differential MOKE effects,
\begin{eqnarray}
    \Delta\theta_k &=& \theta_k^{FM:TI}-\theta_k^{SkX:TI}\\
    \Delta R_{av} &=& R_{av}^{FM:TI}-R_{av}^{SkX:TI}
\end{eqnarray}
where $R_{av}=(R_++R_-)/2$ is the average reflectivity.
We take $\arctan[\tan(\theta_k)]$ to avoid Kerr rotations greater than $\pm90$ deg.

The phase diagram for both the Kerr rotation and the average reflectivity is
shown in Fig. \ref{fig:3D} (a) and (c) for the N\'{e}el type skyrmion.
Both N\'{e}el- and Bloch-type skyrmions have identical phase diagrams
where $\theta_{k}$ and $R_{av}$ vary periodically with $\omega$ and $d_1$.
There are several periodic parts of this phase diagram that are insensitive to errors in film-thickness or spectral tuning.
By tuning $d_1$ a large Kerr rotation can be obtained for any frequency $\gtrsim 100$ meV.
The reflectivity would, however, be low in this case when $\theta_k$ is maximized.
The differential $\Delta \theta_{k}$ and $\Delta R_{av}$ effects are shown in
Fig. \ref{fig:3D} (b) and (d), respectively.
Generally $\Delta R_{av}$ is low.
The FM:TI and SkX:TI have slightly different periodicity.
Therefore the differential Kerr rotation can reach up to $\pm90^\circ$ depending on $d_1$.
The phase diagram also suggests that either the N\'{e}el- or Bloch-type skyrmions could be used for such a magneto-optic recording device.

%It is shown that there are several parts of the optical spectra that can be made relatively insensitive to errors in film-thickness or spectral tuning.
%It is seen that $d_1$ plays a crucial role is maximizing the differential MOKE.
%The maximum differential Kerr rotation always occurs periodically in $d_1$ depending on $\omega$.
%The film thickness $d_1$ bellow 2 $\mu$m does not make much of a difference as long as the device operates between 20-25 meV.
%The differential Kerr rotation can reach upto $\pm90$ deg, the sign of which depends on $d_1$.
%The phase diagram also suggests that either the N\'{e}el- or Bloch-type skyrmions could be used for such a MO recording device.

\begin{figure}
\centering
\includegraphics[width=1\columnwidth]{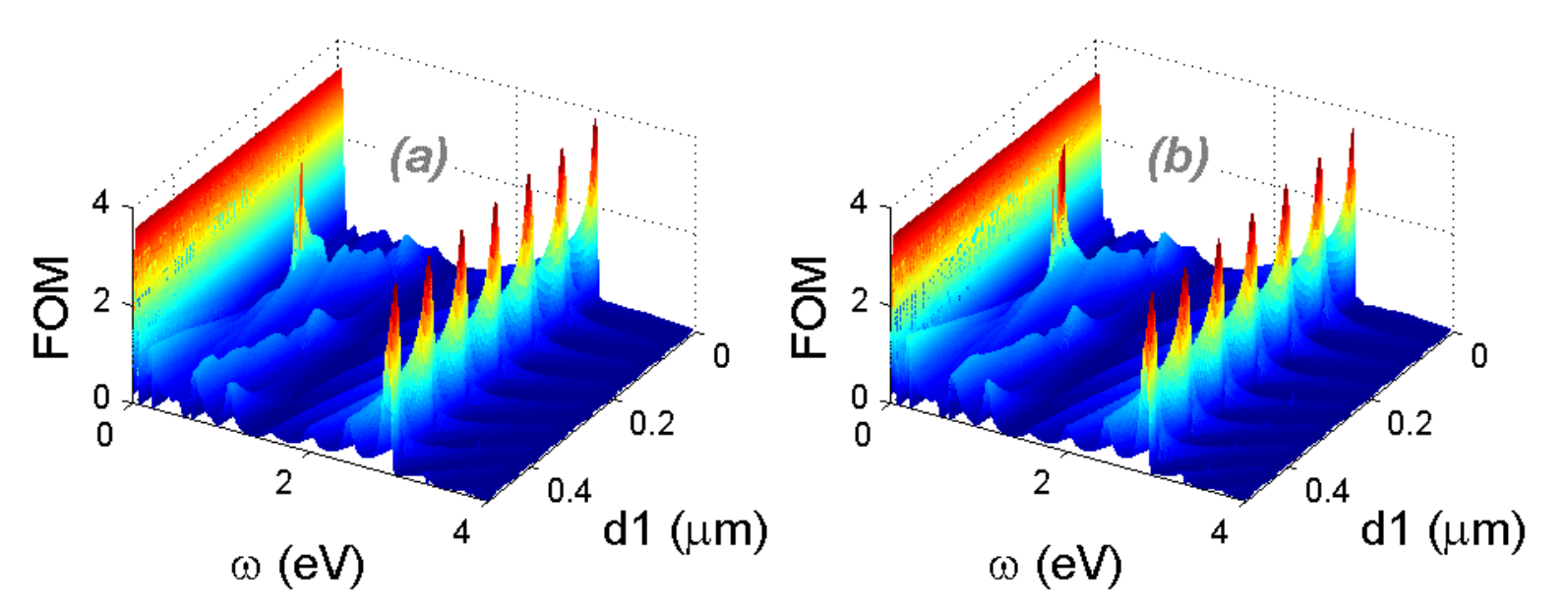}
\caption{Figure of merit for (a) FM:TI (b) N\'{e}el SkX:TI}
\label{fig:FOM}
\end{figure}

%================== FOM ==============================================================================
Typically, large Kerr rotations are accompanied by low reflectivities which reduces the effectiveness of a MO device.
The overall useful MO signal can be quantified by the figure of merit (FOM)\cite{Mansuripur1986,Challener1995,Taussig2008prb} for the Kerr rotation configuration, defined here as,
\begin{eqnarray}
% \nonumber % Remove numbering (before each equation)
  FOM &=& |\theta_k| \times \max(|r_+|,|r_-|).
  \label{FOM}
\end{eqnarray}
This can be used to characterize MO memory and other MO devices such as modulators and isolators.

For our device, the FOM is shown in Fig. \ref{fig:FOM} as a function of $\omega$ and $d_1$.
At $\omega\approx\omega_p\approx~2.8$ eV, the FOM peaks, arising from the free electron behavior,
periodically as a function of $d_1$.
The FOM resonance at $\omega\approx~0.52$ eV, and $d_1=0.2 ~\mu$m is due to the features shown in Fig. \ref{fig:TFres}.
Even though $\theta_k\sim 90^o$, the FOM is not the highest due to low $R_\pm$.
The FOM peak here is higher for the N\'{e}el SkX:TI than it is for the FM:TI.
Also note that the N\'{e}el and Bloch SkX:TI's FOM are identical.

Finally, the most fascinating result is that the FOM at $\omega\approx38$ meV is  independent of $d_1$.
This is an encouraging result for experiments and for devices as this implies
that the MO-FOM is independent of any error in the \CS film-thickness.
Also, this FOM at the gap energy is also the highest in the entire FOM phase diagram.

%%==============================================================================
%\subsection{Fermi Level Dependence}
%%===============================================================================

\begin{figure}
\centering
\includegraphics[width=1\columnwidth]{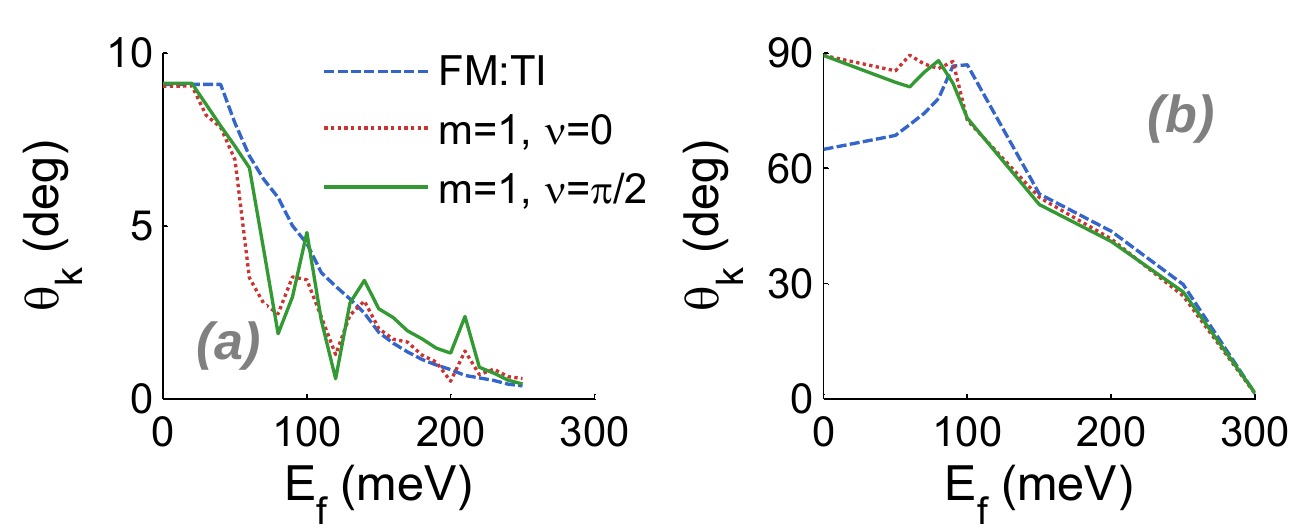}
\caption{(a) Low energy spectral regime Kerr rotation, $\theta_k$,  as a function of Fermi energy, $E_f$, calculated around $\omega\sim$ 38 meV, which is where $\theta_k^{max}$ occurs in Fig. \ref{fig:SIlow} at $E_f=0$.
(b) Thin-film thickness ($d_1=200$ nm) induced Kerr rotation in the intermediate spectral regime, as a function of Fermi energy. Kerr rotations are calculated around $\omega\sim$ 520 meV, which is where $\theta_k^{max}$ occurs for $E_f=0$ in Fig. \ref{fig:TFres}.}
\label{fig:Ef}
\end{figure}

Finally, we study  electro-optic switching effects as a function of Fermi energy.
In Fig. \ref{fig:Ef}(a), $\theta_k$ is shown as a function of Fermi energy $E_f$.
All values are shown at the $\omega$ at which $\theta_k^{max}$ occurs for $E_f=0$ in the low energy regime (see Fig. \ref{fig:SIlow}).
Similarly Fig. \ref{fig:Ef}(b), $\theta_k$ is shown as a function of $E_f$ in the intermediate energy regime.
Here $\omega$ for each magnetic texture is where the resonances
induced by the \CS film thickness occur in Fig. \ref{fig:TFres}(a).

In Fig. \ref{fig:Ef}(a), all three textures show distinct behavior.
While $\theta_k^{max}$ for the FM:TI shows a monotonic decrease with increasing $E_f$,
the N\'{e}el type and Bloch type SkXs show somewhat oscillatory non-monotonic behavior.
This can be explained as follows.
In the present model for the FM:TI all the states above the gap are degenerate and free electron like.
Hence $\theta_k$ decreases monotonically with $E_f$.
Whereas in case of the SkX:TI, the states just above the gap are non-degenerate and the bands are split.
%
%Hence there are $k$-points at which there are no states as $E_f$ moves up in energy.
%
This leads to additional peaks and valleys in the DOS, which causes the sudden jumps in Fig. \ref{fig:Ef}(a) as $E_f$ moves up in energy.
Lastly $\theta_k$ as a function of $E_f$ differs for the N\'{e}el type and Bloch type SkXs because their density of states is different above the gap as shown in Fig. \ref{fig:band}.
These effects can be numerically heightened if $J_H$ is increased tenfold.
The band-splittings in Fig. \ref{fig:band} for the two SkXs, then evolve into energy gaps where the SkXs have different Chern numbers in the gaps. This will lead to very distinct MOKE signatures as a function of $E_f$.

%This effect manifests itself in the MOKE signatures,
%even though there are only high energy pseudo-gaps for the current value of $J_H$.

These $E_f$ dependent non-monotonic $\theta_k$ effects are not seen in Fig. \ref{fig:Ef}(b),
where the MOKE resonances are induced by the \CS thin-film.
Here, the N\'{e}el type and Bloch type SkXs are indistinguishable,
but they both differ from the FM:TI.
$\theta_k$ at $E_f=0$ is much smaller for the FM:TI because $d_1= 200$ nm is not optimal in this case.
The FM:TI and SkX:TI show distinct differences only for low $E_f$.

%TI:SkX interfaces as a function of Fermi energy, $E_f$ does not affect the Bloch type Skrymion's
%MOKE much till the Fermi energy crosses over into the higher continuum bands.
%While for the N\'{e}el type skyrmion an immediate noticeable jump in the maximum Kerr rotation occurs almost as soon as $E_f$ is changed due
%to its low-energy DOS being more continuum like. Overall this makes the N\'{e}el type SkX:TI much more appealing for a electro-optic device.

%==============================================
%\section{Summary}
%==============================================
In summary, the MOKE from SkX-hosting thin-film B20 type compounds interfaced with
TI structures displays rich physics with important device applications.
%
%in the SKX:TI and FM:TI systems. and  magneto-optic recording and other electro-optic devices.
%the MOKE signatures arising from an SkX lattice formed at the interface of a thin-film B20 compound and a semi-infinite TI are
%closely examined in this paper.
%Our purpose is twofold.
%First, we wish to see if in anyway the MOKE signatures can provide useful information about the texture and topology of the TI:SkX system.
%Second, we want to see how useful the TI:SkX system is for magneto-optic recording and other electro-optic devices.
%
%For this purpose
High FOM is obtained from the low energy topological MOKE, the thin-film induced enhancement of MOKE,
and the MOKE occurring at the high energy plasma frequency.
The MOKE-FOM phase diagram shows that the low energy peaks below the TI's bulk energy gap are large and independent of $d_1$.
For the thin-film induced resonance,
the differential MOKE can be  large for the FM and SkX states, which is useful for device applications.
As $E_{f}$ is swept above the exchange gap of the surface state, $\theta_k$ decreases monotonically  for FM:TIs,
and it is distinctly oscillatory and non-monotonic for SkX:TIs.
These distinguishing $\theta_k(E_f)$ features are not seen for the thin-film induced MOKE resonance.
With a large FOM, optical readout can lead to high density and high fidelity MO memory devices and electro-optic devices such as modulators and optical-isolators\cite{Ma2013,Onbasli2016}.

%=====================================================================================
\emph{Acknowledgement:}
This work was supported by the NSF ECCS-1408168 Physical Mechanisms and Limits of Skyrmions for
Information Processing and Storage and as part of the
Spins and Heat in Nanoscale Electronic Systems (SHINES)
an Energy Frontier Research Center funded by the U.S. Department of Energy,
Office of Science, Basic Energy Sciences under Award \#DE-SC0012670.
%
%\bibliography{bibliography,apssamp}

\end{document}